# Terahertz spin conductance probes of coherent and incoherent spin tunneling through MgO tunnel junctions

R. Rouzegar[1,2], M.A. Wahada[3,4], A.L. Chekhov[1,2], W. Hoppe[4], J. Jechumtál[5], L. Nádvorník[5], M. Wolf[2], T.S. Seifert[1], S.S.P. Parkin[3], G. Woltersdorf[4], P.W. Brouwer[1], T. Kampfrath[1,2]

1. Department of Physics, Freie Universität Berlin, 14195 Berlin, Germany
2. Department of Physical Chemistry, Fritz Haber Institute of the Max Planck Society, 14195 Berlin, Germany
3. Max Planck Institute for Microstructure Physics, Weinberg 2, 06120 Halle, Germany
4. Institut für Physik, Martin-Luther-Universität Halle, 06120 Halle, Germany
5. Faculty of Mathematics and Physics, Charles University, Ke Karlovu 3, 121 16 Prague, Czech Republic

**Abstract.**

We study femtosecond spin currents through MgO tunneling barriers in CoFeB(2 nm)|MgO($d$)|Pt(2 nm) stacks by terahertz emission spectroscopy. To obtain transport information independent of extrinsic experimental factors, we determine the complex-valued spin conductance $\tilde{G}_d(\omega)$ of the MgO layer (thickness $d$ = 0-6 Å) over a wide frequency range ($\omega/2\pi$ = 0.5-8 THz). In the time ($t$) domain, $G_d(t)$ has an instantaneous and delayed component that point to (i) spin transport through Pt pinholes in MgO, (ii) coherent spin tunneling and (iii) incoherent resonant spin tunneling mediated by defect states in MgO. A remarkable signature of (iii) is its relaxation time that grows monotonically with $d$ to as much as 270 fs at $d$ = 6 Å, in full agreement with an analytical model. Our results indicate that terahertz spin conductance spectroscopy will yield new and relevant insights into ultrafast spin transport for a wide range of materials.

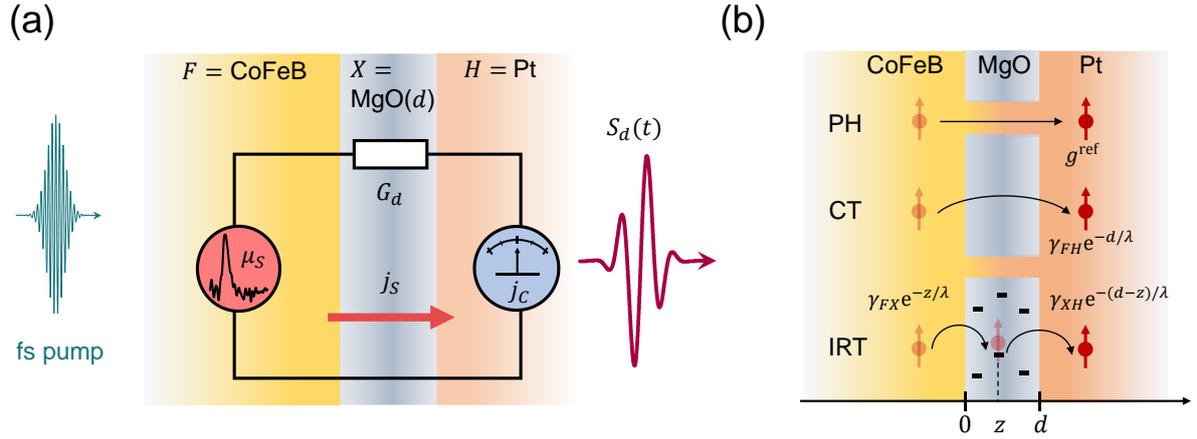

Fig. 1. Spin conductance spectroscopy of a layer $X$. (a) A $F|X|H$ stack is excited by a femtosecond laser pulse. The resulting transient spin voltage $\mu_s(t)$ in the ferromagnetic layer $F$ (here CoFeB) drives a spin current $j_S$ through $X$ (here MgO). The spin current arriving in the heavy metal layer $H$ (here Pt) is converted into an in-plane charge current $j_C$ that is measured by detecting the THz pulse it emits. The dynamics of $\mu_s(t)$ is determined using a sample with $d=0$, which exhibits a spectrally flat interface spin conductance $g^{\mathrm{ref}}$. (b) Considered spin-transport channels through MgO: flow through conducting pinholes (PH), coherent tunneling (CT) and incoherent resonant tunneling (IRT). For IRT, $\gamma_{FX}\mathrm{e}^{-z/\lambda}$ is the probability of an electron to tunnel from $F=$ CoFeB to a defect at position $z$ in $X=$ MgO, whereas $\gamma_{XH}\mathrm{e}^{-(d-z)/\lambda}$ refers to the tunneling probability from $z$ to $H=$ Pt.

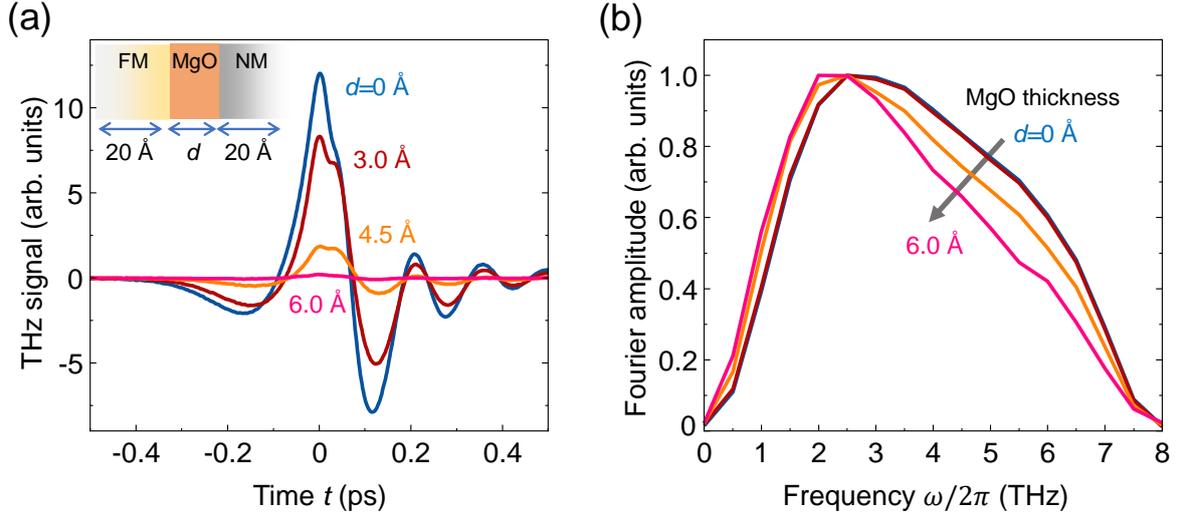

Fig. 2. THz-emission raw data from CoFeB(2 nm)|MgO($d$)|Pt(2 nm). (a) Time-domain electro-optic THz signals $S_d(t)$ odd in the CoFeB magnetization $\boldsymbol{M}$ for various MgO thicknesses of $d = 0$, 3.0, 4.5 and 6.0 Å. The signal for $d = 0$ is the reference signal $S^{\mathrm{ref}} = S|_{d=0}$. (b) Fourier amplitude spectra normalized to their maximum of the signals shown in panel (a).

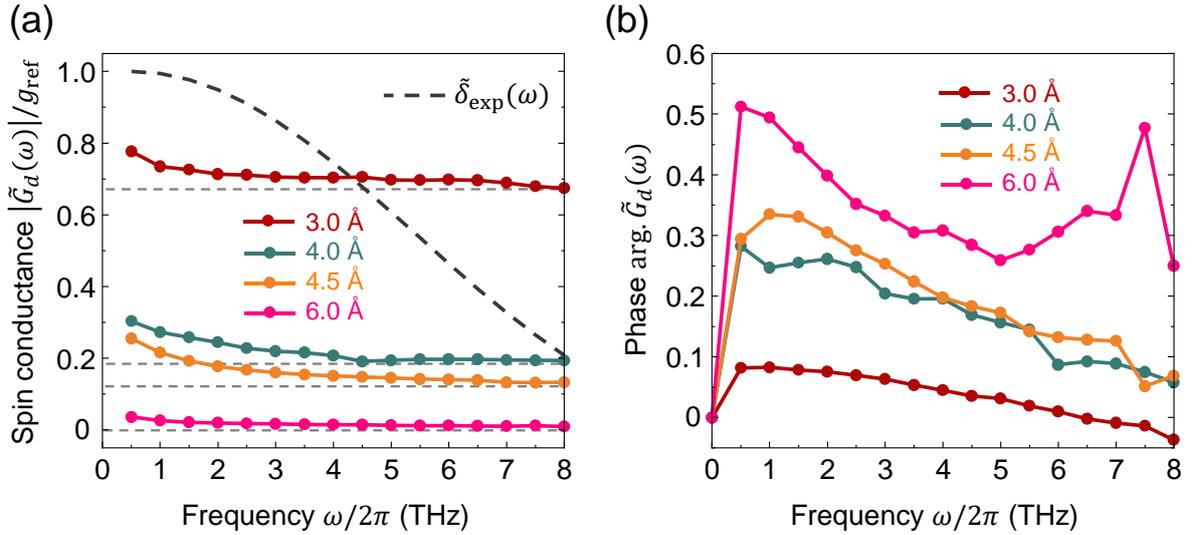

Fig. 3. Frequency-domain THz spin conductance $\tilde{G}_d(\omega)$ of MgO layers with various thicknesses $d$. (a) Fourier amplitude $|\tilde{G}_d(\omega)/g^{\mathrm{ref}}|$ vs frequency $\omega/2\pi$, where $g^{\mathrm{ref}} = \tilde{G}_{d=0}(\omega)$ is spectrally flat. The dashed line shows the Fourier transformation $\tilde{\delta}_{\mathrm{exp}}(\omega)$ of the experimental $\delta$-function. (b) Spectral phase $\arg \tilde{G}_d(\omega)$.

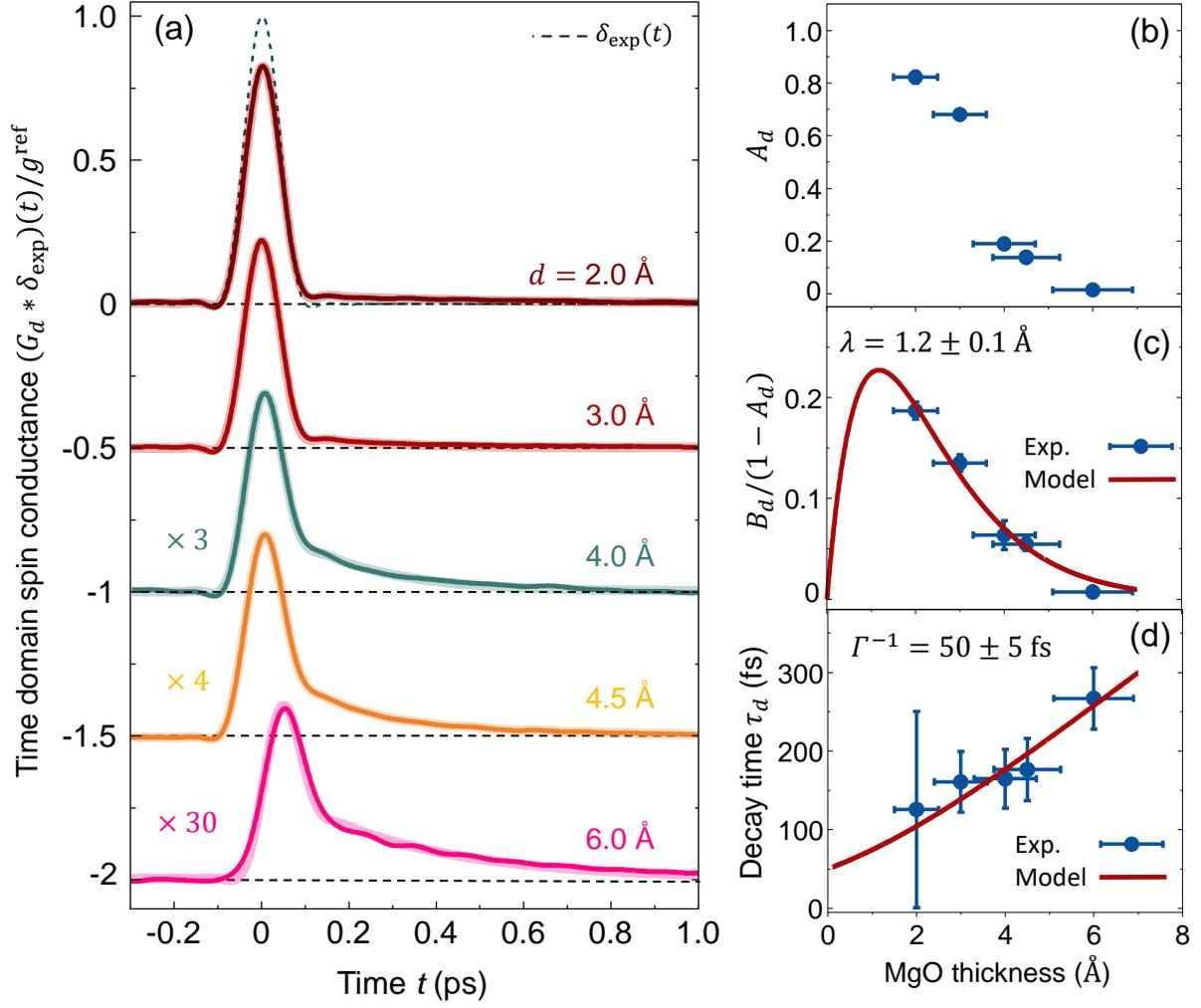

Fig. 4. THz spin transport through MgO layers. (a) Measured time-domain THz spin conductance of MgO films for various thicknesses $d$ (thin solid lines). The spin conductance $G_d(t)$ can be understood as the current that is driven through the MgO layer by a $\delta(t)$-like spin voltage in the CoFeB layer. Due to the finite time resolution, the $\delta$-function is broadened to the fictitious impulsive spin voltage $\delta_{\exp}(t)$ (dashed black line). Accordingly, the extracted conductivity curves are also broadened and display $G_d * \delta_{\exp}$. Thick solid lines are fits based on Eq. (3). (b) Amplitude $A_d \approx f_d^{\mathrm{PH}}$ of the instantaneous, i.e., $\delta_{\exp}(t)$-like component of $G_d(t)$ vs MgO thickness $d$ (blue circles). (c) Amplitude $B_d/(1-A_d) \approx g_{d,0}^{\mathrm{IRT}}/g^{\mathrm{ref}}$ of the delayed current component vs MgO thickness (blue circles). The red solid line is a fit based on Eq. (5) with $\lambda = (1.2 \pm 0.1)$ Å. (d) Decay time $\tau_d$ of the IRT current component vs MgO thickness (blue circles). The red solid line is a fit based on Eq. (5) with $\Gamma^{-1} = (50 \pm 5)$ fs.

**Motivation.** Transport of electron spin angular momentum is of central importance for future spintronic devices. To keep pace with other information carriers such as electrons in field-effect transistors [1] and photons in optical fibers [2], spintronic transport needs to be pushed to terahertz (THz) bandwidth and, thus, femtosecond time scales. Interesting applications of ultrafast spin transport include emission of broadband THz pulses [3] and generation of spin torque [4] for ultrafast magnetization switching.

To better understand and optimize THz spin transport, we need tools to adequately characterize it. An important quantity is the spin conductance $G = I_S/\Delta\mu_S$, which quantifies how much spin current $I_S$ is obtained when a spin-voltage drop $\Delta\mu_S$ is applied across a conductor [Fig. 1(a)]. We focus on longitudinal spin transport, which can be described by populations of spin-up and spin-down electron states. At frequencies $\ll 1$ THz, spin transport measurements typically rely on contacts [5-7]. At THz frequencies, however, measurement procedures of $G$ still need to be developed.

In this work, we introduce an approach to measure the spin conductance of a thin film $X$ between a ferromagnetic metal layer $F$ and a heavy-metal layer $H$ [Fig. 1(a)]. Using THz emission spectroscopy [8], we obtain the complex-valued spin conductance $\tilde{G}(\omega)$ at frequencies $\omega/2\pi = 0.5\text{-}8$ THz. In the time domain, $G(t)$ vs time $t$ can simply be understood as the spin current that would be obtained for a $\delta(t)$-like spin-voltage pulse. Our procedure is demonstrated for the ubiquitous tunnel-barrier material $X = $ MgO and reveals dynamic signatures of coherent and incoherent spin tunneling. We expect that THz spin conductance spectroscopy will provide important insights into ultrafast spin transport in a wide range of materials.

**THz spin-conductance spectroscopy.** The general idea of THz spin-conductance spectroscopy is shown in Fig.1. It relies on a $F|X|H$ stack, where the ferromagnetic layer $F$ (here CoFeB) serves as spin-current source, $X$ is the layer under investigation, and the heavy-metal layer $H$ (here Pt) acts as detector. First, a femtosecond laser pulse induces an ultrafast generalized spin voltage $\mu_S(t)$ in $F = $ CoFeB [9,10] that drives a spin current through $X = $ MgO. Second, the spin current $j_S(t)$ arriving in the $F = $ Pt detection layer is converted into an in-plane charge current $j_C(t) \propto j_S(t)$ by the inverse spin Hall effect (ISHE). Third, $j_C(t)$ gives rise to the emission of a measurable ultrashort THz electromagnetic pulse with electric field $E(t)$ [3,11].

In the frequency domain, the spin conductance of MgO is given by an Ohm-type law, $\tilde{j}_S(\omega) = \tilde{G}_d(\omega)\tilde{\mu}_S(\omega)$, where $d$ is the thickness of $X$. The spin voltage in Pt is neglected owing to the short spin lifetime in Pt [12]. To determine $\tilde{\mu}_S(\omega)$, we conduct a reference measurement on a sample without interlayer (i.e., $d = 0$), where $\tilde{j}_S^{\text{ref}}(\omega) = \tilde{G}^{\text{ref}}(\omega)\mu_S(\omega)$ with known $\tilde{G}^{\text{ref}}$.

In our experiment, we measure electrooptic signals $S(t)$ and $S^{\text{ref}}(t)$ rather than the emitted THz electric field $E(t)$. However, $S$ and $E$ are related by an instrument response function that cancels in the frequency domain when the reference measurement is considered [13]. As derived in Supplemental Note 1, the $X$ spin conductance $\tilde{G}_d$ normalized to the frequency-independent reference conductance $\tilde{G}^{\text{ref}}$ is fully determined by the observables $\tilde{S}$ and $\tilde{S}^{\text{ref}}$ through

$$\frac{\tilde{G}_d(\omega)}{g^{\text{ref}}} = \frac{\tilde{j}_S(\omega)}{\tilde{j}_S^{\text{ref}}(\omega)} = \frac{\tilde{S}(\omega)}{\tilde{S}^{\text{ref}}(\omega)}. \quad (1)$$

In the last step of Eq. (1), we took advantage of the facts that (i) the measured optical absorptance and THz impedance of all the samples of our experiment are the same and, thus, cancel, and that (ii) the spin conductance $\tilde{G}^{\text{ref}}(\omega) = g^{\text{ref}}$ of the $F|$Pt interface is constant over the frequency interval considered here (see Supplemental Note 2).

**Experimental details.** As spin conductor $X$, the nonmagnetic insulator MgO is chosen. It is extensively used as a tunnel barrier in magnetic tunnel junctions, which provide large tunneling magnetoresistance for applications in nonvolatile memory devices [14-17].

The samples are sub||TaN(1.5 nm)|CoFeB(2 nm)|MgO($d$)|Pt(2 nm)|TaN(1.5 nm) stacks with various MgO thicknesses of $0 \leq d \leq 15$ Å. The sample with $d = 0$ is the reference sample. All layers are grown on an MgO substrate (sub) by DC magnetron sputtering in ultrahigh vacuum at a base pressure of $4 \times 10^{-3}$ mbar, except the MgO layer, which is grown in the same vacuum chamber by radio-frequency sputtering using an off-axis gun tilted 90° from the substrate plane [18]. Atomic force microscopy shows a roughness <2 Å of all layers [18].

In the THz emission experiments [Fig. 1(a)], the $F$ magnetization $M$ of $F$ = CoFeB is saturated by an external magnetic field of about 10 mT. The sample is excited by a train of linearly polarized ultrashort laser pulses (central wavelength 790 nm, nominal pulse duration 10 fs, pulse energy 2 nJ, repetition rate 80 MHz) from a Ti:Sapphire laser oscillator, focused to a spot of 30 µm full width of half maximum (FWHM) of intensity. The emitted THz pulse is detected by electro-optic sampling in a GaP(110) crystal (thickness 250 µm) using linearly polarized probe pulses (0.6 nJ) from the same laser. This procedure yields the THz signal $S(t)$ as a function of delay $t$ between probe and THz pulse [10].

**THz signals.** Figure 2(a) and Fig. S1 show THz emission signals $S(t)$ odd in $M$ from CoFeB|MgO|Pt stacks for various MgO thicknesses $d$. Signal components even in $M$ are minor.

The signal for $d = 0$ is known to be dominated by ultrafast spin transport from CoFeB into Pt with $\tilde{G}_{d=0}(\omega) = g^{\text{ref}} = \text{const}_\omega$ [9-11,19]. When $d$ is increased from 0 Å to 6.0 Å, the signal amplitude decreases by a factor of 60 [Fig. 2(a)]. One can show that, in this range, the THz signal is dominated by spin transport from $F$ through MgO into $H$ and the ISHE in $H$, consistent with the signal origin for $d = 0$ (see Supplemental Material).

Interestingly, with increasing $d$, the signal amplitude not only drops, but the initially sharp temporal features become smoother [Fig. 2(a)]. This trend suggests the emergence of slower components in the spin-current dynamics $j_S(t)$. It is confirmed by the normalized Fourier spectra [Fig. 2(b)], whose width decreases with increasing $d$ and whose maximum shifts to lower frequencies.

**Frequency-domain spin conductance.** To gain more intrinsic insight into the spin transport through MgO from the signals of Fig. 2, we make use of Eq. (1) to determine the normalized spin conductance $\tilde{G}_d(\omega)/g^{\text{ref}}$ as a function of frequency $\omega/2\pi = 0.5$-8 THz. The modulus $|\tilde{G}_d(\omega)/g^{\text{ref}}|$ is displayed in Fig. 3(a). It consists of a frequency-independent component and an additional contribution below 3 THz. As expected from Fig. 2(a), we observe a drastic overall amplitude reduction with increasing $d$. At the same time, spectral weight is shifted to frequencies below 3 THz. The spectral phase $\arg \tilde{G}_d(\omega)$ [Fig. 3(b)] of the spin conductance and its slope vs $\omega$ increase with $d$, indicating an increasingly non-instantaneous response.

**Time-domain spin conductance.** To obtain the time-domain spin conductance $G_d(t)$, we inversely Fourier-transform $\tilde{G}_d(\omega)$. The resulting normalized $G_d(t)/g^{\text{ref}}$ is shown in Fig. 4(a) for all MgO thicknesses.

Note that $G_d(t)$ can be considered as the spin current through MgO that is induced by an impulsive spin voltage. Therefore, $G_d(t)$ for $d = 0$ would ideally equal a $\delta(t)$-type signal. However, as the bandwidth of our setup is finite, the extracted $G_d(t)$ is the response to a broadened $\delta$-like pulse $\delta_{\text{exp}}(t)$. To achieve a unipolar $\delta_{\text{exp}}(t)$, which is as short as possible and free of oscillations, we apply a suitable window function $\tilde{\delta}_{\text{exp}}(\omega)$ to $\tilde{G}_d(\omega)/g^{\text{ref}}$ prior to Fourier transformation [20]. Our $\tilde{\delta}_{\text{exp}}(\omega)$ [dashed line in Fig. 3(a)] approaches zero at 12 THz. The resulting $\delta(t)$-type spin-voltage pulse $\delta_{\text{exp}}(t)$ has a FWHM of 90 fs [dashed line in Fig. 4(a)] and defines the time resolution of $G_d(t)$.

For $d > 0$, the extracted $G_d(t)$ traces [Fig. 4(a)] consist of an initial instantaneous response with a shape similar to $\delta_{\text{exp}}(t)$ plus a subsequently decaying component, whose relative weight increases with $d$. This result is fully consistent with the spectra of Fig. 3(a), where the broad background and feature below 3 THz correspond to the instantaneous and slower time-domain response [Fig. 4(a)].

**Interpretation.** Thin MgO films have been studied extensively in the past. For $d < 10$ Å, structural imperfections are reported, in particular oxygen vacancies [21-26] and pinholes connecting $F$ and $H$ [27,28]. Accordingly, we consider 3 possible contributions to the total spin current through MgO in

CoFeB|MgO|Pt stacks [Fig. 1(b)]: (i) spin transport through Pt- or CoFeB-filled pinholes (PH) in the MgO film [28,29], (ii) coherent off-resonant electron tunneling (CT) through the MgO tunnel barrier [30-32] and (iii) incoherent resonant tunneling (IRT) through intermediate defect states in the vicinity of the Fermi energy of the CoFeB and Pt layer [23,26,33-35]. As the currents (i)-(iii) add up independently [Fig. 1(b)], the spin conductance is the sum

$$G_d = G_d^{\mathrm{PH}} + G_d^{\mathrm{CT}} + G_d^{\mathrm{IRT}}. \qquad (2)$$

While the PH and CT processes are instantaneous [10,36] on the time scale of our experimental resolution of 90 fs [dashed black line in Fig. 4(a)], the IRT mechanism may require more time to proceed.

We briefly discuss each mechanism in more detail. Regarding $G^{\mathrm{PH}}$, one expects that, for $d < 6$ Å, MgO grows in islands [27] or exhibits pinholes [28]. The pinholes are filled with Pt of the subsequently grown Pt layer and, thus, provide a conductive channel between the CoFeB and Pt layer [Fig. 1(b)]. $G_d^{\mathrm{PH}}$ depends on the in-plane areal fraction $f_d^{\mathrm{PH}}$ of pinholes according to $G_d^{\mathrm{PH}} \propto g^{\mathrm{ref}} f_d^{\mathrm{PH}}$. Thus, for $f_d^{\mathrm{PH}} = 1$, the reference situation ($d = 0$) is recovered. Because the PH contribution is instantaneous analogous to the reference sample, we obtain $G_d^{\mathrm{PH}}(t) = f_d^{\mathrm{PH}} g^{\mathrm{ref}} \delta(t)$.

CT [Fig. 1(b)] through the entire MgO barrier is instantaneous on the scale of our time resolution [36]. Therefore, $G_d^{\mathrm{CT}}(t) = g_d^{\mathrm{CT}}(1 - f_d^{\mathrm{PH}})\delta(t)$, where the coefficient $g_d^{\mathrm{CT}}$ is the amplitude of the impulsive spin current in the absence of pinholes.

Regarding IRT, oxygen vacancies are known to provide localized electronic states within the MgO band gap and, thus, open up a resonant transport channel [21,24,25]. In the simplest IRT realization, an electron tunnels from $F$ into a MgO vacancy and, subsequently, into $H$ [Fig. 1(b)], similar to resonant tunneling in quantum wells [37-40]. One can quantify the resonant tunneling as $G_d^{\mathrm{IRT}}(t) = g_d^{\mathrm{IRT}}(t)(1 - f_d^{\mathrm{PH}})$, where $g_d^{\mathrm{IRT}}(t)$ is the IRT-related spin conductance of an MgO barrier without pinholes. Based on Fig. 1(b), we model $g_d^{\mathrm{IRT}}(t)$ by a single-sided exponential decay $g_{d,0}^{\mathrm{IRT}} \Theta(t) e^{-t/\tau_d}$, where $\Theta(t)$ is the Heaviside step function and $\tau_d$ can be considered as the characteristic time of IRT.

With these specifications, Eq. (2) turns into

$$\frac{G_d(t)}{g^{\mathrm{ref}}} = A_d \delta(t) + B_d \Theta(t) e^{-t/\tau_d}, \qquad (3)$$

where

$$A_d = f_d^{\mathrm{PH}} + \left(1 - f_d^{\mathrm{PH}}\right) \frac{g_d^{\mathrm{CT}}}{g^{\mathrm{ref}}}, \qquad B_d = \left(1 - f_d^{\mathrm{PH}}\right) \frac{g_{d,0}^{\mathrm{IRT}}}{g^{\mathrm{ref}}}. \qquad (4)$$

We fit the convolution of $G_d$ [Eq. (3)] and our time resolution $\delta_{\mathrm{exp}}$ to the time-domain data of Fig. 4(a), where $\tau_d$, $A_d$, $B_d$ and possible deviations $t_0$ from time zero [i.e., $t \to t - t_0$ in Eq. (3)] due to substrate thickness variations [41] are fit parameters (see Supplemental Material).

The model fits [Fig. 4(a)] describe our data excellently. The resulting amplitude $A_d$ of the $\delta(t)$-like contribution [Fig. 4(b)] decreases monotonically with MgO thickness $d$. One can show that $A_d \approx f_d^{\mathrm{PH}}$ (see Supplemental Material). This assignment and Fig. 4(b) are consistent with previous work in which $f_d^{\mathrm{PH}}$ was found to decrease with increasing MgO thickness $d$ on a scale of 2 Å [28]. According to calculations [18], MgO layers close to the CoFeB and Pt interface are slightly metallic. They have a total thickness of about 4 Å and can be understood to have a large PH fraction. This effect may explain the pronounced drop of $A_d$ from $d = 3.0$ Å to 4.0 Å [Fig. 4(b)].

Using $A_d \approx f_d^{\mathrm{PH}}$, we can now determine $g_{d,0}^{\mathrm{IRT}}/g^{\mathrm{ref}} \approx B_d/(1 - A_d)$, i.e., the peak amplitude of the IRT contribution [Eq. (3)], which is found to decrease strongly with increasing $d$ [Fig. 4(c)]. Remarkably, the characteristic IRT time $\tau_d$ is found to grow with $d$ [Fig. 4(d)], consistent with the increase of the slope of $\arg \tilde{G}_d(\omega)$ vs $\omega$ [Fig. 3(b)].

**Model of dynamic IRT.** Qualitatively, we suggest the following dynamic scenario for the IRT conductance $g_d^{\mathrm{IRT}}(t) = g_{d,0}^{\mathrm{IRT}} \Theta(t) e^{-t/\tau_d}$. At time $t = 0$, a $\delta(t)$-like spin voltage in CoFeB drives instantaneous tunneling of spin-polarized electrons from CoFeB to MgO defect states. Simultaneously, spin-unpolarized electrons from Pt are transferred to maintain local charge neutrality.

The subsequent tunneling events from occupied defect states to CoFeB or Pt are stochastic. Their rate $\partial_t N^\sigma$ is proportional to the number $N^\sigma$ of spins $\sigma = \uparrow$ or $\downarrow$ occupying the defect state. Accordingly, we obtain a (i) simple temporal exponential decay of $N^\sigma$ and, thus, the $\sigma$-current from defects to Pt. (ii) The characteristic time of this process increases with increasing $d$ because the tunneling from defects to Pt becomes less likely as $d$ and, thus, the average distance of defect and Pt grow. Finally, as two tunneling events over distance $d$ are involved, (iii) the spin current from CoFeB to Pt decreases with increasing $d$. The implications (i), (ii), (iii) of our model are fully consistent with the experimental results of Figs. 4(a), 4(c) and 4(d), respectively.

More quantitatively, a rate-equation treatment and the assumption of vanishing out-of-plane charge current (see Supplemental Material) reproduce the relationship $g_d^{\mathrm{IRT}}(t) = g_{d,0}^{\mathrm{IRT}} \Theta(t) e^{-t/\tau_d}$ with

$$g_{d,0}^{\mathrm{IRT}} \propto d e^{-d/\lambda}, \qquad \tau_d^{-1} = \Gamma \frac{1 - e^{-d/\lambda}}{d/\lambda}. \tag{5}$$

Here, $\Gamma^{-1}$ quantifies the spin lifetime in a defect state for an infinitely thin MgO layer, and $\lambda$ is the spin decay length in a MgO barrier. Eq. (5) provides good fits to the data of Fig. 4(b), (c) and yields $\Gamma^{-1} = (50 \pm 5)$ fs and $\lambda = (1.2 \pm 0.1)$ Å in good agreement with theoretical predictions [31,42].

**Discussion.** Measurement of the THz spin conductance relies on the measurement of just two THz emission signals, without having to know any instrument response functions, in contrast to extraction of the charge source current. This approach can be extended to, in principle, any layer $X$ other than MgO, if the following conditions are fulfilled: (1) The signal exclusively arises from the spin current $j_S$ arriving in $H$. (2) $j_S$ solely originates from the spin voltage $\mu_S$. (3) The presence of the layers $X$ and $H|X$ does not change the $\mu_S$ dynamics of $F$.

For our CoFeB|MgO($d$)|Pt system, (1) is fulfilled for the reference sample ($d = 0$), but also for $d \neq 0$ because the ISHE of Pt dominates spin-to-charge conversion of the metal stack (see Supplemental Material). (2) is fulfilled for the reference sample [10] and, thus, for $d \neq 0$, too. (3) is fulfilled because the presence of $X$ in $F|H$ stacks was shown to leave the spin-voltage dynamics unchanged [10]. If $X$ has a large optical and THz refractive index, the modified pump-propagation and current-to-THz-field conversion need to be accounted for (see Supplemental Material).

In conclusion, we demonstrate THz conductance spectroscopy for the example of MgO barriers. In the time domain, we find ultrafast signatures of IRT, i.e., an exponentially decaying spin current with a relaxation time that increases with the MgO layer thickness. This behavior arises because the tunneling probability decreases with increasing thickness of the tunneling barrier. We anticipate that our method can be used to measure the spin conductance of a large set of materials, ranging from simple metals [43] to complex materials such as antiferromagnets [44-46], potentially also involving a transverse spin-current component.